\documentclass[12pt]{iopart}
\newcommand{\bdiv}{\bnabla \cdot}
\newcommand{\brot}{\bnabla \times}
\newcommand{\brn}{\bi{r}^{n}}
\newcommand{\brp}{\bi{r}^{\prime}}
\newcommand{\Dn}{\mathcal{D}_n}
\newcommand{\intv}{\int_V \! \rmd v \,}
\newcommand{\intvv}[1]{\langle \, #1 \, \rangle _V}
\newcommand{\ints}{\int_S \! \rmd s \,}
\newcommand{\intss}[1]{\langle \, #1 \, \rangle _S}
\newcommand{\intomp}{\int \! \rmd \Omega^{\prime} \,}
\newcommand{\mrs}{\rm s}
\newcommand{\mrv}{\rm v}
\newcommand{\mrL}{\rm L}
\newcommand{\mrT}{\rm T}
\newcommand{\LHS}{\rm LHS}
\newcommand{\RHS}{\rm RHS}
\newcommand{\Rn}{\bi{R}^{(n)}}
\usepackage{iopams}
\usepackage{graphicx, txfonts}

\begin{document}

\title[]{Generalized integral formulation of electromagnetic Cartesian multipole moments
\footnote[1]{This is an Author's Original Manuscript of an article whose final and definitive form, the Version of Record, has been published in the {\it Journal of Electromagnetic Waves and Applications} (published online 16 July 2013) [copyright Taylor \& Francis], available online at: http://www.tandfonline.com/10.1080/09205071.2013.819473.}
}

\author{J. Niitsuma\footnote[7]{email: niitsuma@jaist.ac.jp}}

\address{School of Materials Science, Japan Advanced Institute of Science and Technology,
1-1 Asahidai, Nomi, Ishikawa 923-1292, Japan}

\begin{abstract}
We study integral expressions of electromagnetic multipole moments of arbitrary order in Cartesian coordinates. The volume and surface integrals of charge-induced and current-induced multipole moment tensors are formulated and the relationship between them is discussed. Full surface integral expressions for the multipole moment are also obtained. We further extend the formulation to introduce another kind of dipole moment, which is similar to the charge-induced and current-induced multipole moments and is found in a vector decomposition formula. 
\end{abstract}

\maketitle

\section{Introduction}

Multipoles emerge in the calculation of various quantities in electromagnetism, e.g., force, torque, energy, radiation and interaction~\cite{Jackson1}--\cite{Vrejoiu2}. This significant and useful concept is a necessary consequence of a series expansion of a function selected according to the quantity to be calculated. The most common and often dominant multipole is the dipole, and in some cases higher-order multipoles are considered.

Multipoles have two sources: electric charge and electric current. For an electrically neutral material with a volume $V$ bounded by a closed surface $S$ and placed in vacuum, the sources of multipole moments are volume and surface densities of polarization charges $\rho_{\mrv}$ and $\rho_{\mrs}$, and those of magnetization currents $\bi{j}_{\mrv}$ and $\bi{j}_{\mrs}$. If they are independent of time, we write them in SI units as~\cite{Jackson1, Kovetz1, Leung1, Martin1}
\begin{eqnarray}
	&\rho_{\mrv} = -\bdiv \, \bi{P}, \qquad &\rho_{\mrs} = \bi{n} \cdot \bi{P}, \label{echarge} \\
	&\bi{j}_{\mrv} = \brot \bi{M}, \qquad &\bi{j}_{\mrs} = - \bi{n} \times \bi{M}, \label{ecurrent}
\end{eqnarray}
where $\bi{P}$ and $\bi{M}$ are the effective electric and magnetic polarization density vectors, respectively, and $\bi{n}$ is the outward unit vector normal to $S$. The surface charge and current densities in (\ref{echarge}) and (\ref{ecurrent}) should always be present in a real material~\cite{Martin1} because it is usually bounded by a surface. If $\bi{P}$ ($\bi{M}$) is a constant nonzero vector inside $V$ and is zero outside $V$, the total dipole moment will vanish without the surface term because $\rho_{\mrv}=0$ ($\bi{j}_{\mrv}=\boldsymbol{0}$) in the entire space.

It is not a simple task, especially for undergraduate students, to understand the physical meaning of (\ref{echarge}) and (\ref{ecurrent}). One way to clarify it is to calculate the integral of dipole moment vectors generated from $\rho_{\mrv/\mrs}$ and $\bi{j}_{\mrv/\mrs}$. Using some vector formulas together with the divergence theorem, we confirm the following~\cite{Kovetz1,Leung1}:
\begin{eqnarray}
	\intv \bi{r} \rho_{\mrv} + \ints \bi{r} \rho_{\mrs} = \intv \bi{P}, \label{pe} \\*
	\frac{1}{2} \intv \bi{r} \times \bi{j}_{\mrv} + \frac{1}{2} \ints \bi{r} \times \bi{j}_{\mrs} = \intv \bi{M}, \label{me}
\end{eqnarray}
where $\bi{r}$ is the position vector, and $\intv$ and $\ints$ are volume and surface integrals, respectively.

Vector identities (\ref{pe}) and (\ref{me}) indicate that (i) the integrated electric (magnetic) dipole moment generated from the polarization charge (magnetization current) is correctly given by the volume integral of the electric (magnetic) polarization density vector, and (ii) if we formally set $\bi{P}=\bi{M}$, the left-hand sides of (\ref{pe}) and (\ref{me}) are identical despite their different expressions.

A question arises as to whether (i) and (ii) can be generalized to multipole moments of any order. Dubovik and Tosunyan proved that (ii) is true even for a higher-order multipole moment, but they did not consider surface contributions~\cite{Dubovik1}. In the same article there was no mention of (i), probably because multipole moments were expressed by spherical harmonics. To understand the physical meaning of multipole moments, we need to represent them in terms of Cartesian coordinates~\cite{Jackson1, Cipriani1, Applequist1, Vrejoiu1, Vrejoiu2}.

In this work, we generalize (\ref{pe}) and (\ref{me}), and therefore (i) and (ii) to a higher-order multipole moment in Cartesian coordinates. The volume and surface integrals of charge-induced and current-induced multipole tensors are calculated first for the symmetric tensor case and then for the symmetric traceless tensor case. Next, we present full surface integral formulations for the volume integral of a multipole moment. Finally, we extend the formulation to introduce another kind of dipole moment generated by an angular momentum operator and discuss its similarity to the charge-induced and current-induced dipole moments. We show that the angular momentum-induced dipole emerges in some vector decomposition formula. This article is accessible to undergraduate students with a fundamental knowledge in vector and tensor calculus~\cite{Arfken1}.

\section{Symmetric tensor case}

\subsection{Definitions}
We assume a general effective polarization density vector field $\bi{Q}$ that is a continuous and differentiable vector function of $\bi{r}$ and is independent of time. $\bi{Q} \neq \boldsymbol{0}$ only in a  volume $V$ bounded by a closed surface $S$. The volume densities of the generalized polarization charge and current are defined as $\rho_{\mrv} = -\bdiv \, \bi{Q}$ and $ \bi{j}_{\mrv} = \brot \bi{Q}$, respectively. The corresponding generalized polarization surface charge and current densities are $\rho_{\mrs} = \bi{n} \cdot \bi{Q}$ and $ \bi{j}_{\mrs} = -\bi{n} \times \bi{Q}$, respectively. We assume that there are no free charges and currents in the system studied here.

The $n$th ($2^n$-pole) Cartesian moment of volume/surface {\it charge} density with respect to the origin in $V$ is defined as $\brn \rho_{\mrv/\mrs} = \bi{r}^{n-1} \bi{q}_{1\mrv/\mrs}$, where $\bi{q}_{1\mrv/\mrs}= \bi{r} \rho_{\mrv/\mrs}$ is the volume/surface charge-induced dipole density, and $\brn = \overbrace{\bi{r} \bi{r} \cdots \bi{r} }^n$ is a Cartesian tensor of rank $n$ ($\geq 1$)~\cite{Applequist1}. On the other hand, the $n$th Cartesian moment of volume/surface {\it current} density is defined as $\brn \times \bi{j}_{\mrv/\mrs} = \bi{r}^{n-1} \bi{q}_{2\mrv/\mrs}$~\cite{Vrejoiu2}, where $\bi{q}_{2\mrv/\mrs}=\bi{r} \times \bi{j}_{\mrv/\mrs}$ is termed the volume/surface current-induced dipole density. (The common current-induced dipole density is $\frac{1}{2} \, \bi{q}_{2\mrv/\mrs}=\frac{1}{2} \, \bi{r} \times \bi{j}_{\mrv/\mrs}$.)

A tensor is symmetric if its components are invariant under an interchange of any pair of their indices. Therefore, the charge moment density tensor $\bi{r}^{n-1} \bi{q}_{1\mrv/\mrs}$ is symmetric, whereas the current moment density tensor $\bi{r}^{n-1} \bi{q}_{2\mrv/\mrs}$ is not. However, an asymmetric moment tensor $\bi{r}^{n-1} \bi{a}$, where $\bi{a}$ is any vector, can be symmetrized as follows:
\begin{eqnarray}
	\bi{r}^{n-1} \bi{a} \, \to \, \bi{r}^{n-1} \bi{a} + \bi{r}^{n-2} \bi{a} \bi{r} + \cdots + \bi{a} \bi{r}^{n-1} = \bi{a} \cdot \bnabla \brn. \label{sym}
\end{eqnarray}
In component form, the right-hand side of (\ref{sym}) is calculated in the following way, where $x_i$ represents a Cartesian component of the vector $\bi{r}=(x_1, x_2, x_3)=(x,y,z)$, $\partial_i=\partial/\partial x_i$, $\partial_i x_j = \delta_{ij}$ (the Kronecker delta) and the repeated subscript implies summation from 1 to 3:
\begin{eqnarray}
	\bi{a} \cdot \bnabla \brn &= a_i \partial_i (x_{i_1} x_{i_2} \cdots x_{i_n}) \nonumber \\*
	&= a_i (\delta_{ii_1} x_{i_2} \cdots x_{i_n} \! + x_{i_1} \delta_{ii_2} \cdots x_{i_n} \! + x_{i_1} x_{i_2} \cdots \delta_{ii_n}) \nonumber \\*
	&= a_{i_1} x_{i_2} \cdots x_{i_n} \! + x_{i_1} a_{i_2} \cdots x_{i_n} \! + x_{i_1} x_{i_2} \cdots a_{i_n}. \nonumber
\end{eqnarray}
This is a symmetric $(n-1)$th moment tensor of $\bi{a}$. From (\ref{sym}), the symmetric $n$th moment tensor of current density is written as $\bi{q}_{2\mrv/\mrs} \cdot \bnabla \brn$. The ($n$-time) moment tensor of charge density, which is symmetric itself, is rewritten similar to (\ref{sym}) by using $\bi{r} \cdot \bnabla \brn = n \brn$: 
\begin{eqnarray}
	n \bi{r}^{n-1} \bi{q}_{1\mrv/\mrs} = \bi{q}_{1\mrv/\mrs} \cdot \bnabla \brn. \nonumber
\end{eqnarray}

Let us consider the following differential identity to calculate multipole Cartesian moment integrals:
\begin{eqnarray}
	\partial_i(Q_j x_k \partial_l \brn) = (\partial_i Q_j) x_k \partial_l \brn + Q_j \partial_i(x_k \partial_l \brn). \label{master}
\end{eqnarray}
By virtue of the divergence theorem, the volume integral of (\ref{master}) yields
\begin{eqnarray}
	\intv (-\partial_i Q_j) x_k \partial_l \brn + \ints n_i Q_j x_k \partial_l \brn = \intv Q_j \partial_i(x_k \partial_l \brn), \nonumber
\end{eqnarray}
or simply
\begin{eqnarray}
	\intvv{(-\partial_i Q_j) x_k \partial_l \brn} + \intss{n_i Q_j x_k \partial_l \brn} = \intvv{Q_j \partial_i(x_k \partial_l \brn)}. \label{masterint}
\end{eqnarray}

\subsection{Charge-induced multipole}
We take contractions between $i$ and $j$, $k$ and $l$ in (\ref{masterint}) or apply $\delta_{ij} \delta_{kl}$ to (\ref{masterint}) and take the sum over $i$ and $k$. In this case, the left- and right-hand sides of (\ref{masterint}) become
\begin{eqnarray}
	\LHS &= \intvv{ (-\partial_i Q_i) x_k \partial_k \brn} + \intss{ n_i Q_i x_k \partial_k \brn} \nonumber \\*
	&= \intvv{ \bi{q}_{1\mrv} \! \cdot \bnabla \brn} + \intss{ \bi{q}_{1\mrs} \! \cdot \bnabla \brn}, \label{1L}
\end{eqnarray}
\begin{eqnarray}
	\RHS &= \intvv{ Q_i \partial_i(x_k \partial_k \brn)} = n \intvv{ \bi{Q} \cdot \bnabla \brn}. \label{1R}
\end{eqnarray}
Again, $x_k \partial_k \brn=n \brn$ is used in (\ref{1R}). From (\ref{1L}) and (\ref{1R}), we obtain the expression for the $n$th moment tensor of the polarization charge:
\begin{eqnarray}
	n^{-1} \left[ \strut \intvv{ \bi{q}_{1\mrv} \! \cdot \bnabla \brn} + \intss{ \bi{q}_{1\mrs} \! \cdot \bnabla \brn} \right] = \intvv{ \bi{Q} \cdot \bnabla \brn}. \label{1fin} 
\end{eqnarray}
This is quite similar to (\ref{pe}), which is actually derived from (\ref{1fin}) for $n=1$.

\subsection{Current-induced multipole}
Next, we apply $-\varepsilon_{lkm} \varepsilon_{mij}= \delta_{ik} \delta_{jl}-\delta_{il} \delta_{jk}$ to (\ref{masterint}), where $\varepsilon_{lkm}$ is a standard anti-symmetric unit tensor of rank 3.
\begin{eqnarray}
	\LHS &= \intvv{ \varepsilon_{lkm} x_k \varepsilon_{mij} (\partial_i Q_j) \partial_l \brn} + \intss{ \varepsilon_{lkm} x_k \varepsilon_{mij} (-n_i) Q_j \partial_l \brn} \nonumber \\*
	&= \intvv{ \bi{q}_{2\mrv} \! \cdot \bnabla \brn} + \intss{ \bi{q}_{2\mrs} \! \cdot \bnabla \brn}. \label{2L} 
\end{eqnarray}
On the other hand, the integrand on the right-hand side becomes
\begin{eqnarray}
	 (\delta_{ik} \delta_{jl}-\delta_{il} \delta_{jk}) \, Q_j \partial_i(x_k \partial_l \brn) = Q_j \partial_i(x_i \partial_j \brn) - Q_j \partial_i(x_j \partial_i \brn). \nonumber
\end{eqnarray}
The first term is rewritten as follows: 
\begin{eqnarray}
	Q_j \partial_i (x_i \partial_j \brn) &= Q_j \partial_i [\partial_j (x_i \brn) - \delta_{ij} \brn] \nonumber \\*
	&= Q_j [\partial_j (\delta_{ii} \brn + x_i \partial_i \brn) - \delta_{ij} \partial_i \brn ] \nonumber \\*
	&= Q_j \partial_j (3 \brn + n \brn - \brn) = (n+2) Q_j \partial_j \brn. \nonumber 
\end{eqnarray}
For the second term, we have
\begin{eqnarray}
	Q_j \partial_i(x_j \partial_i \brn) = Q_j (\delta_{ij} \partial_i \brn + x_j \partial_i \partial_i \brn) = Q_j (\partial_j \brn + x_j \partial_i \partial_i \brn). \label{iljk} 
\end{eqnarray}
Therefore, it follows that
\begin{eqnarray}
	\RHS &= (n+1) \intvv{ Q_j \partial_j \brn} - \intvv{ Q_j x_j \partial_i \partial_i \brn} \nonumber \\*
	&= (n+1) \intvv{ \bi{Q} \cdot \bnabla \brn} - \intvv{ \bi{Q} \cdot \bi{r} \triangle \brn}, \label{2R}
\end{eqnarray}
where $\triangle=\partial_i \partial_i$ denotes the Laplacian. From (\ref{2L}) and (\ref{2R}), we find the following expression for the $n$th moment tensor of the polarization current:
\begin{eqnarray}
	(n+1)^{-1} & \left[ \strut \intvv{ \bi{q}_{2\mrv} \! \cdot \bnabla \brn} + \intss{ \bi{q}_{2\mrs} \! \cdot \bnabla \brn} \right] \nonumber \\*
	&= \intvv{ \bi{Q} \cdot \bnabla \brn} - (n+1)^{-1} \intvv{ \bi{Q} \cdot \bi{r} \triangle \brn}. \label{2fin} 
\end{eqnarray}
Except for the last term including $\triangle \brn$, this is similar to (\ref{me}), which is recovered by setting $n=1$ in (\ref{2fin}).

\section{Symmetric traceless tensor case}

Here, we consider transforming (\ref{1fin}) and (\ref{2fin}) into traceless tensors. A symmetric tensor $\bi{A} ^{(n)}$ of rank $n$ ($\geq 1$) is rendered traceless by the detracer operator $\Dn$ defined in~\cite{Applequist1} as follows:
\begin{eqnarray}
	\Dn A_{\alpha_1 \cdots \alpha_n}^{(n)} =& \frac{1}{(2n-1)!!} \sum_{m=0}^{[n/2]} (-1)^m (2n-2m-1)!! \nonumber \\*
	& \times \sum_{T \{ \alpha \} }\delta_{\alpha_1 \alpha_2} \cdots \delta_{\alpha_{2m-1} \alpha_{2m}} A_{\nu_1 \nu_1 \cdots \nu_m \nu_m \alpha_{2m+1} \cdots \alpha_{n}}^{(n)}, \label{detracer}
\end{eqnarray}
where $[n/2]$ denotes the integer part of $n/2$ and the sum over $T \{ \alpha \}$ is the sum over all permutations of the symbols $\alpha_1, \cdots, \alpha_n $ that yield distinct terms.

One can see from (\ref{detracer}) that the operator $\Dn$ generates a linear combination of terms that are derived from the original tensor $\bi{A} ^{(n)}$ by replacing and contracting the indices $\alpha_1, \cdots, \alpha_n $. We note that  integral and scalar differential operators do not affect the sequence of the indices of a tensor. These two facts show that $\Dn$ commutes with the operators $\intvv{\cdots}$, $\intss{\cdots}$, $\bi{a} \cdot \bnabla$ and $\triangle$ in (\ref{1fin}) and (\ref{2fin}).

Hence, applying the detracer to symmetric tensors (\ref{1fin}) and (\ref{2fin}) simply changes the symmetric tensor $\brn$ into the symmetric traceless tensor $\Rn$, which is expressed as~\cite{Cipriani1,Applequist1}
\begin{eqnarray}
	\Rn = \Dn \brn = \frac{(-1)^n}{(2n-1)!!} r^{2n+1} \bnabla^n \frac{1}{r}, \quad r = |\bi{r} |, \nonumber
\end{eqnarray}
with $\bnabla^n = \overbrace{ \bnabla \bnabla \cdots \bnabla }^n$. $\Rn$ is regular at the origin~\cite{Cipriani1}. For example, $\bi{R}^{(1)}=\bi{r}$ and $\bi{R}^{(2)}=x_i x_j-\frac{1}{3}\delta_{ij}r^2$. One can easily show that $\triangle \Rn$ vanishes with the use of $\bi{r} \cdot \! \bnabla^{n+1} r^{-1} = -(n+1) \bnabla^n r^{-1}$~\cite{Applequist1}. Therefore, from (\ref{1fin}) and (\ref{2fin}), we finally obtain the following expression:
\begin{eqnarray}
	\alpha_{i,n} \left[ \strut \intvv{ \bi{q}_{i\mrv} \! \cdot \bnabla \Rn} + \intss{ \bi{q}_{i\mrs} \! \cdot \bnabla \Rn} \right] = \intvv{ \bi{Q} \cdot \bnabla \Rn}, \quad (i=1,2), \label{totfin}
\end{eqnarray}
where $\alpha_{1,n}=n^{-1}$ and $\alpha_{2,n}=(n+1)^{-1}$. This is the general version of (\ref{pe}) and (\ref{me}) and a main result of this article. Two conclusions, similar to (i) and (ii), are drawn from (\ref{totfin}).
\begin{list}{}{}
\item (i') The sum of the volume and surface integrals of the symmetric traceless Cartesian moments of $\bi{q}_{i\mrv}$ and $\bi{q}_{i\mrs}$ is equal to the volume integral of the symmetric traceless Cartesian moment of $\bi{Q}$ of the same order.
\item (ii') The Cartesian multipole moments induced by the polarization charge ($i=1$) and current ($i=2$) are identical after integration for any order. 
\end{list}
Therefore, it was shown that (i) and (ii) are correct for a multipole moment of arbitrary order.

\section{Full surface integral expression for a multipole moment}

According to Helmholtz's theorem~\cite{Arfken1}, the vector $\bi{Q}$ can be written as a sum of two parts, $\bi{Q}=\bi{Q}^{\mrT} + \bi{Q}^{\mrL}$, where $\bi{Q}^{\mrT}$ and $\bi{Q}^{\mrL}$ are the transverse and longitudinal components of $\bi{Q}$, respectively. They satisfy $\bdiv \, \bi{Q}^{\mrT}=0$ and $\brot \bi{Q}^{\mrL}=\boldsymbol{0}$. Replacing $\bi{Q}$ in (\ref{totfin}) by $\bi{Q}^{\mrT}$ for $i=1$ and by $\bi{Q}^{\mrL}$ for $i=2$, we get
\begin{eqnarray}
	\alpha_{1,n} \intss{ \bi{q}_{1\mrs \mrT} \cdot \bnabla \Rn} = \intvv{ \bi{Q}^{\mrT} \cdot \bnabla \Rn}, \label{Tsurfint} \\*
	\alpha_{2,n} \intss{ \bi{q}_{2\mrs \mrL} \cdot \bnabla \Rn} = \intvv{ \bi{Q}^{\mrL} \cdot \bnabla \Rn}, \label{Lsurfint} 
\end{eqnarray}
where $\bi{q}_{1\mrs \mrT}= \bi{r} \bi{n} \cdot \bi{Q}^{\mrT}$ and $\bi{q}_{2\mrs \mrL}=\bi{r} \times (-\bi{n} \times \bi{Q}^{\mrL})$. Adding both sides of (\ref{Tsurfint}) and (\ref{Lsurfint}) yields
\begin{eqnarray}
	\intss{ ( \, \alpha_{1,n} \, \bi{q}_{1\mrs \mrT} + \alpha_{2,n} \, \bi{q}_{2\mrs \mrL} \, ) \cdot \bnabla \Rn} = \intvv{ \bi{Q} \cdot \bnabla \Rn}. \label{surfint1}
\end{eqnarray}
This is a full surface integral expression for the volume integral of a multipole moment.

Here, we introduce the scalar $\phi$ and vector $\bi{A}$ potentials to rewrite (\ref{surfint1}). Because $\bi{Q}=\bi{Q}^{\mrL}+\bi{Q}^{\mrT}=\bnabla \phi+\brot \bi{A}$, the integrand on the right-hand side of (\ref{surfint1}) becomes
\begin{eqnarray}
	Q_i \partial_i \Rn &= [ \, \partial_i \phi + \varepsilon_{ijk} (\partial_j A_k) ] \partial_i \Rn \nonumber \\*
	&= \partial_i \! \left( \phi \partial_i \Rn \right) \! - \phi \partial_i \partial_i \Rn + \varepsilon_{ijk} \! \left[ \partial_j \! \left( A_k \partial_i \Rn \right) - A_k \partial_j \partial_i \Rn \right] \rule{0pt}{2.5ex} \nonumber \\*
	&= \partial_i \! \left( \phi \partial_i \Rn \right) \! + \varepsilon_{ijk} \partial_j \! \left( A_k \partial_i \Rn \right). \nonumber
\end{eqnarray}
Then, volume integration and the divergence theorem yield an alternative full surface integral form:
\begin{eqnarray}
	\intss{ ( \, \bi{n} \phi + \bi{n} \times \bi{A} \, ) \cdot \bnabla \Rn} = \intvv{ \bi{Q} \cdot \bnabla \Rn}. \label{surfint2}
\end{eqnarray}
The same result is also obtained by directly calculating the left-hand side of (\ref{surfint1}) using Stokes' theorem.

Equations (\ref{surfint1}) and (\ref{surfint2}) show that the surface integral of the symmetric traceless Cartesian moment of $\alpha_{1,n} \, \bi{q}_{1\mrs \mrT}+\alpha_{2,n} \, \bi{q}_{2\mrs \mrL}$ or $\bi{n} \phi+\bi{n} \times \bi{A}$ is equivalent to the volume integral of the symmetric traceless Cartesian moment of $\bi{Q}$ of the same order. The advantage of this formula is that only the surface values of a pair, $(\bi{Q}^{\mrT}, \bi{Q}^{\mrL})$ or $(\phi, \bi{A})$, are needed to determine the volume-integrated multipole moment.

\section{Another expression for a multipole moment}

We now consider another type of contraction of (\ref{masterint}) by applying $-\varepsilon_{lmj} \varepsilon_{mki}= \delta_{ij} \delta_{kl}-\delta_{il} \delta_{jk}$ to (\ref{masterint}):
\begin{eqnarray}
	\LHS &= \intvv{ \varepsilon_{lmj} \varepsilon_{mki} (x_k \partial_i Q_j) \partial_l \brn} + \intss{ \varepsilon_{lmj} \varepsilon_{mki} x_k (-n_i) Q_j \partial_l \brn} \nonumber \\*
	&= \intvv{ \bi{q}_{3\mrv} \! \cdot \bnabla \brn} + \intss{ \bi{q}_{3\mrs} \! \cdot \bnabla \brn}, \nonumber
\end{eqnarray}
\begin{eqnarray}
	\RHS &= (\ref{1R}) - \intvv{(\ref{iljk})} = (n-1) \intvv{ \bi{Q} \cdot \bnabla \brn} - \intvv{ \bi{Q} \cdot \bi{r} \triangle \brn}, \nonumber
\end{eqnarray}
where $\bi{q}_{3\mrv}= (\bi{r} \times \bnabla) \times \bi{Q}$ and $\bi{q}_{3\mrs}=[ \bi{r} \times (-\bi{n}) ] \times \bi{Q}$. By implementing the detracer $\Dn$, we find
\begin{eqnarray}
	\intvv{ \bi{q}_{3\mrv}} + \intss{ \bi{q}_{3\mrs}} = \boldsymbol{0}, \quad (n=1) \label{5fin1}
\end{eqnarray}
\begin{eqnarray}
	(n-1)^{-1} &\left[ \strut \intvv{ \bi{q}_{3\mrv} \! \cdot \bnabla \Rn} + \intss{ \bi{q}_{3\mrs} \! \cdot \bnabla \Rn} \right] \nonumber \\*
&= \intvv{ \bi{Q} \cdot \bnabla \Rn}. \quad (n \geq 2) \label{5fin} 
\end{eqnarray}

From (\ref{5fin}), one can see that (\ref{totfin}) is also valid for $i=3$ with $\alpha_{3,n}=(n-1)^{-1}$ ($n \geq 2$). This means that the three dipole densities $\bi{q}_{i\mrv/\mrs}$ ($i=1$--$3$) are equivalent up to multiplication by $\alpha_{i,n}$ as long as the integral multipole operation is considered. Hence, relation (\ref{totfin}) provides alternative ways to integrate charge-induced and current-induced multipole moments by using $\bi{q}_{3\mrv/\mrs}$. The expressions for $\bi{q}_{i\mrv/\mrs}$ and $\alpha_{i,n}$ are summarized in table~\ref{table}.
\begin{table}
\caption{\label{table}Volume and surface densities of three dipole moments and $\alpha_{i,n}$ factors.}
	\begin{center}
		\begin{tabular}{ccccc}
		\br
			$i$ & Dipole type & $\bi{q}_{i\mrv}$ & $\bi{q}_{i\mrs}$ & $\alpha_{i,n}^{-1}$ \\
		\mr
				1 & Charge-induced & $\bi{r} (-\bdiv \bi{Q})$ & $\bi{r} \bi{n} \cdot \bi{Q}$ & $n$ \\
				2 & Current-induced & $\bi{r} \times (\brot \bi{Q})$ & $\bi{r} \times (-\bi{n} \times \bi{Q})$ & $n+1$ \\
				3 & -- & $(\bi{r} \times \bnabla) \times \bi{Q}$ & $[\bi{r} \times (-\bi{n}) ] \times \bi{Q}$ & $n-1$ \\
		\br
		\end{tabular}
	\end{center}
\end{table}

The dipole moment density $\bi{q}_{3\mrv}=\bi{L} \times \bi{Q} $, where $\bi{L} = \bi{r} \times \bnabla$ corresponds to the angular momentum operator, appears in a vector decomposition formula~\cite{Lomont1}--\cite{Moses1}:
\begin{equation}
	\bi{Q}(\bi{r}) = \bi{L} \varphi (\bi{r}) + \bi{L} \times \bi{F} (\bi{r}) + \bi{F} (\bi{r}). \label{Ldecomp}
\end{equation}
The scalar and vector {\it potentials} $\varphi$ and $\bi{F}$ can be written in the following forms (see Appendix):
\begin{eqnarray}
	\varphi (\bi{r}) &= L^{-2} (\bi{L} \cdot \bi{Q}), \label{Ldecomp1} \\*
	\bi{F} (\bi{r}) &= - L^{-2} (\bi{L} \times \bi{Q}), \label{Ldecomp2} \\*
	L^{-2} f &= \frac{1}{4 \pi} \intomp \ln(1-\hat{\bi{r} \rule{0pt}{1.5ex}} \cdot \hat{\brp \rule{0pt}{1.5ex}}) \, f(\bi{r}^{\prime}), \label{Ldecomp3}
\end{eqnarray}
where $L^{-2}$ is the inverse of the differential operator $L^2=\bi{L} \cdot \bi{L}$ and $\intomp$ represents an integration over the entire solid angle~\cite{Wilcox1, Backus1}. As seen in (\ref{Ldecomp2}), $\bi{q}_{3\mrv}=\bi{L} \times \bi{Q}$ is the source of the vector potential $\bi{F}$ and $\bi{q}_{3\mrs}$ may be its surface analogue. Because the integrated total dipole moment is always zero, as shown in (\ref{5fin1}), its dipole nature can be defined only locally.

\section{Conclusions}

We constructed symmetric traceless Cartesian multipole moment tensors induced by polarization charge and current, which are independent of time. We found that generalized multipole moment densities are identical after volume and surface integrations, and that they are also the same as the volume integral of the polarization vector density. Therefore, the generalized version of (\ref{pe}) and (\ref{me}) for a higher-order multipole moment was established. Alternative full surface integral forms for the volume integral of the polarization multipole moment tensor were also obtained.

We introduced another type of dipole moment density vector generated by an angular momentum operator and showed that it is equivalent to the charge-induced and current-induced dipole moment density vectors as long as the integral multipole operation is considered. We found that the angular momentum-induced dipole moment is related to a specific vector potential that appears in a vector decomposition formula.

\ack I would like to thank T Itoh, K Edamatsu, T Sekiguchi, H Yokoyama, M Yoneya and T Shimoda for their support. I also thank J C Terrillon for proofreading the manuscript.

\appendix

\section*{Appendix. Proof of (\ref{Ldecomp})--(\ref{Ldecomp3})}
\setcounter{section}{1}

First, we prove the following identity for an angular momentum operator $\bi{L}=\bi{r} \times \bnabla$ and any differentiable vector function $\bi{Q}$:
\begin{eqnarray}
	L^2 \bi{Q} = \bi{L} (\bi{L} \cdot \bi{Q}) - \bi{L} \times (\bi{L} \times \bi{Q}) - \bi{L} \times \bi{Q}. \label{LL}
\end{eqnarray}
We calculate $\bi{L} \times (\bi{L} \times \bi{Q})$ by using both $\varepsilon_{ijk} \varepsilon_{klm} = \delta_{il} \delta_{jm} - \delta_{im} \delta_{jl}$ and the commutation relation for $\bi{L}$, i.e.\ $L_i L_j -L_j L_i = -\varepsilon_{ijn} L_n$:
\begin{eqnarray} 
	\varepsilon_{ijk} L_j \varepsilon_{klm} L_l Q_m &= L_j L_i Q_j - L_j L_j Q_i \nonumber \\*
	&= L_i L_j Q_j + \varepsilon_{ijn} L_n Q_j - L_j L_j Q_i, \nonumber	
\end{eqnarray}
from which (\ref{LL}) follows immediately.

Next, let us consider the equation $L^2 \bi{Q} = \bi{f}$. This has a solution if and only if~\cite{Wilcox1, Backus1}
\begin{eqnarray}
	\int_{S_{\! b}} \! \rmd S \, \bi{f} = \boldsymbol{0}, \label{Lcond}
\end{eqnarray}
where $S_{\! b}$ denotes the spherical surface of radius $b$ centered at the origin. In this case, there is a unique solution $\bi{Q}$ satisfying $\int_{S_{\! b}} \! \bi{Q} =\boldsymbol{0}$, and it is given by $\bi{Q}=L^{-2} \bi{f}$~\cite{Wilcox1, Backus1}. We now verify that (\ref{LL}) satisfies condition (\ref{Lcond}). By using $\bi{L}=b\bi{n} \times \bnabla$ on $S_{\! b}$ and a variant of Stokes' theorem~\cite{Arfken1}, we obtain
\begin{eqnarray}
	\int_{S_{\! b}} \! \rmd S \, \bi{L} = b \int \! \rmd \bi{c}. \label{Stokes}
\end{eqnarray}
The contour integral operator $\int \! \rmd \bi{c}$ is zero for the closed surface $S_{\! b}$. Because all three terms on the right-hand side of (\ref{LL}) begin with the operator $\bi{L}$, their surface integrals $\int_{S_{\! b}} \! \rmd S$ vanish due to (\ref{Stokes}). This shows that (\ref{LL}) satisfies condition (\ref{Lcond}), and therefore has solution $\bi{Q}$.

We now obtain the solution $\bi{Q}$ for (\ref{LL}). Because $L^{-2}$ and $\bi{L}$ commute with each other (as will be shown later),
\begin{eqnarray}
	\bi{Q} = L^{-2} L^2 \bi{Q} &= L^{-2} \left[ \bi{L} (\bi{L} \cdot \bi{Q}) - \bi{L} \times (\bi{L} \times \bi{Q}) - \bi{L} \times \bi{Q} \right] \nonumber \\*
	&= \bi{L} L^{-2} (\bi{L} \cdot \bi{Q}) - \bi{L} \times L^{-2} (\bi{L} \times \bi{Q})  - L^{-2} (\bi{L} \times \bi{Q}), \label{L2inv}
\end{eqnarray}
which completes the proof of (\ref{Ldecomp})--(\ref{Ldecomp3}). This is a unique solution satisfying $\int_{S_{\! b}} \! \bi{Q}=\boldsymbol{0}$.
 
\begin{figure}
	\begin{center}
		\includegraphics[width=4cm,clip]{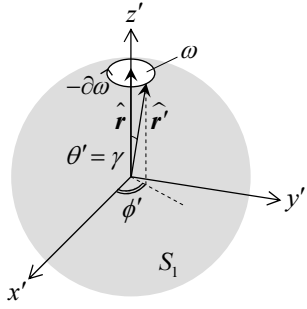}
		\caption{\label{figB1}Unit sphere $S_{\! 1}$ as an integration region to compute (\ref{intc}).}
	\end{center}
\end{figure}

The commutativity between $L^{-2}$ and $\bi{L}$ indicates that $\bi{L} L^{-2} f = L^{-2} \bi{L} f$ for a function $f$ that is differentiable on the unit sphere $S_1$. We rewrite the left-hand side as follows:
\begin{eqnarray}
	\bi{L} L^{-2} f = \bi{L} \intomp h f(\brp) = \intomp (\bi{L} h) f(\brp), \label{Lcomm} 
\end{eqnarray}
where $h=(4 \pi)^{-1}\log(1-\hat{\bi{r} \rule{0pt}{1.5ex}} \cdot \hat{\brp \rule{0pt}{1.5ex}})$. By using $L_i h=-L_i^{\prime} h$ ($\bi{L}^{\prime}=\bi{r}^{\prime} \times \bnabla^{\prime}$), the integrand in (\ref{Lcomm}) becomes
\begin{eqnarray}
	(L_i h) f = -(L_i^{\, \prime} h) f = -L_i^{\, \prime} (hf) + h L_i^{\, \prime} f. \label{LLp} 
\end{eqnarray}
To proceed with the integration of $L_i^{\, \prime} (hf)$ in (\ref{LLp}), we need to consider the singularity of $h$ at $\hat{\brp \rule{0pt}{1.5ex}}=\hat{\bi{r} \rule{0pt}{1.5ex}}$. We first eliminate a small solid angle $\omega$ around the singularity from $S_1$, then take the limit $\omega \to 0$ after the integration (figure~\ref{figB1}).
The surface integral turns into a contour integral in a manner similar to (\ref{Stokes}).
\begin{eqnarray}
		-\int_{S_{\! 1}-\omega} \! \rmd \Omega^{\prime} \, \bi{L}^{\prime} \left[ \, h f(\brp) \, \right] = -\int_{-\partial \omega} \! \rmd \bi{c}^{\prime} h f(\brp). \label{intc}
\end{eqnarray}
To calculate the right-hand side of (\ref{intc}), we select spherical coordinates $(r^{\prime}, \theta^{\prime}, \phi^{\prime})$ with $\bi{r}$ fixed along the $z^{\prime}$-axis, and then $h$ is singular at $\hat{\brp} = \hat{\bi{z}^{\prime}}$. We define $S_{\! 1}$ as $r^{\prime}=1$, $0 \leq \theta^{\prime} \leq \pi$ and $0 \leq \phi^{\prime} < 2 \pi$. Let $\omega$ be a circular region determined by $0 \leq \theta^{\prime} < \gamma$ and $0 \leq \phi^{\prime} < 2 \pi$ on $S_{\! 1}$, as shown in figure~\ref{figB1}. In this case, the contour is $\rmd \bi{c}^{\prime} = - \sin \gamma \, \rmd \phi^{\prime} \hat{\bphi^{\prime}}$ ($0 \leq \phi^{\prime} < 2 \pi$). If $|f| \leq M$ on $S_{\! 1}$ for some positive constant $M$, the contour integral in (\ref{intc}) is estimated as follows:
\begin{eqnarray}
	\left| \, \int_{0}^{2\pi} \! \rmd \phi^{\prime} \hat{\bphi^{\prime}} \sin \gamma \, \ln(1-\cos \gamma) \, f \, \right| \, \leq 2 \pi M \left| \, \sin \gamma \, \ln(1-\cos \gamma) \, \right|, \nonumber 
\end{eqnarray}
and approaches zero as $\gamma \to 0$. Therefore, only the last term survives in (\ref{LLp}). Finally, from (\ref{Lcomm}) and (\ref{LLp}), we obtain
\begin{eqnarray}
	\bi{L} L^{-2} f = \intomp h \, \bi{L}^{\prime} f(\brp) = L^{-2} \bi{L} f, \nonumber 
\end{eqnarray}
i.e.\ the commutativity between $L^{-2}$ and $\bi{L}$ for a function $f$ that is differentiable and bounded on the unit sphere. The functions $\bi{L} \cdot \bi{Q}$ and $\bi{L} \times \bi{Q}$ in (\ref{L2inv}) must also have the same properties to allow the operation in (\ref{L2inv}).

\end{document}